\documentclass[pra,aps,amsmath,amssymb,amsfonts,twocolumn,nofootinbib,longbibliography,superscriptaddress]{revtex4-2}
\usepackage{amssymb}
\usepackage{bm,mathrsfs}
\usepackage{graphicx}
\usepackage{subfigure}
\usepackage{epsfig}
\usepackage{amsmath,bbm}
\usepackage{amsfonts,amssymb}
\usepackage{times}
\usepackage{verbatim}
\usepackage{diagbox}
\usepackage[sort&compress]{natbib}
\usepackage{amsmath}
\usepackage[colorlinks=true,citecolor=blue,linkcolor=blue,urlcolor=blue]{hyperref}
\usepackage[usenames]{color}

\usepackage{float}
\begin{document}

\preprint{APS/123-QED}

\title{Many-body localization properties of fully frustrated Heisenberg spin-1/2 ladder model with next-nearest-neighbor interaction}

\author{Jiameng Hong}
\affiliation{Center for Quantum Sciences and School of Physics, Northeast Normal University, Changchun 130024, China}

\author{Taotao Hu}
\email{hutt262@nenu.edu.cn}
\affiliation{Center for Quantum Sciences and School of Physics, Northeast Normal University, Changchun 130024, China}

\begin{abstract}
Many-body localization (MBL) is an intriguing physical phenomenon that arises from the interplay of interaction and disorder, allowing quantum systems to prevent thermalization. In this study, we investigate the MBL properties of the fully frustrated Heisenberg spin-1/2 ladder model with next-nearest-neighbor hopping interaction along the leg direction and compare it with the Heisenberg spin-1/2 single-chain model with next-nearest-neighbor hopping interaction. We explore the MBL transition using random matrix theory and study the characteristics of entanglement entropy and its variance. Our results show that for the single-chain model, the critical point $w _{1} \sim$ 7.5 ± 0.5, whereas for the frustrated ladder model, $w _{2} \sim$ 10.5 ± 0.5. Moreover, we observe the existence of a many-body mobility edge in the frustrated ladder model.
We also investigate the dynamical properties of the frustrated ladder model and identify the logarithmic growth of entanglement entropy, high fidelity of initial information, and magnetic localization phenomenon in the localized phase.
Finally, we explore the finite-size scaling of the two models. Our findings suggest that interpreting MBL transition as a continuous second-order phase transition yields a better scaling solution than the Kosterlitz-Thouless type transition for our two models, and this difference is more pronounced in the frustrated ladder model compared with the single-chain model.

\end{abstract}

\maketitle

\section{\label{sec:level1}Introduction}
The eigenstate thermalization hypothesis (ETH) \cite{PhysRevE.50.888,PhysRevA.43.2046,doi:10.1080/00018732.2016.1198134} suggests that over long periods of time, all microscopic states of the system are visited with equal probability, i.e. a single many-body eigenstate possesses thermodynamic observables. However, the phenomenon of many-body localization (MBL) \cite{PhysRevLett.95.206603,PhysRevB.77.064426,PhysRevB.82.174411,ALET2018498,RevModPhys.91.021001, RevModPhys.91.021001} challenges the ETH by hindering the approach to equilibrium in strongly disordered interacting quantum systems. This is attributed to the emergence of local integrals of motion (LIOM) \cite{PhysRevB.90.174202,ROS2015420,PhysRevX.7.021018,PhysRevB.97.064204,PhysRevB.97.060201}, which enables the system to retain information about its initial state, impeding the dissemination of information \cite{PhysRevLett.117.040601,RevModPhys.93.025003} and decelerating the diffusion of entanglement \cite{PhysRevB.94.214206,PhysRevLett.110.260601}. Therefore, this is also why MBL is considered a potential candidate for constructing quantum memories. Moreover, this transition depends on the position of the energy spectrum known as the mobility edge, causing mid-spectrum eigenstates to localize at higher disorder values than the edge eigenstates.

Experiments on MBL have also been realized on artificial platforms such as ultracold atoms \cite{PhysRevA.92.041601,PhysRevX.11.021021,PhysRevLett.120.110601}, ion traps \cite{PhysRevA.93.063602,PhysRevApplied.11.011002}, and superconducting qubits \cite{PhysRevB.100.134504,PhysRevLett.120.050507}. Additionally, researchers are also exploring the existence of many-body localization in real materials. Furthermore, the MBL properties in one-dimensional spin systems \cite{PhysRevB.101.014201,PhysRevB.87.134202,PhysRevResearch.3.L032030}, bosonic systems \cite{PhysRevB.108.054201,PhysRevA.100.013628}, and fermionic systems \cite{PRXQuantum.4.040348,PhysRevLett.128.146601}, based on Heisenberg and Ising interactions, have achieved consistency between theory and experiments. In addition, significant progress has been made in the study of many-body localization based on the phenomenological renormalization group \cite{PhysRevLett.121.206601,PhysRevX.5.031033,PhysRevX.5.031032,PhysRevLett.119.110604}. This provides an important reference value for distinguishing the universal classes of many-body localization.

The Heisenberg spin-1/2  ladder model \cite{PhysRevB.53.R2934,PhysRevB.54.13009,PhysRevB.47.3196} has recently attracted great interest. Excitingly, in the experimental realm, H. Yamaguchi et al. have realized ladder models with exchange interactions exhibiting both ferromagnetic and antiferromagnetic behavior \cite{PhysRevLett.73.2626,PhysRevLett.110.157205,PhysRevB.89.220402}. Then, in theoretical aspects, Zheng-Hang Sun et al. have investigated the MBL transition and dynamic properties of the ladder model \cite{PhysRevResearch.2.013163}.

More recently, the frustrated (with crossed interactions) Heisenberg spin ladder model \cite{10.21468/SciPostPhys.11.4.074} has attracted significant attention among researchers. Dominik Hahn et al. have explored the dynamics of the model through out-of-time-ordered correlators (OTOC) and entanglement entropy. Sk Saniur Rahaman and Rojas studied the quantum phases and thermodynamic properties of frustrated spin-1/2 ladders with alternate Ising-Heisenberg exchange interactions, providing a phase diagram for this model \cite{PhysRevB.92.184421}. In contrast to their work, our study focuses more on the many-body localization properties of this model in the presence of disorder.

 The frustrated spin ladder model, often described as a `` quasi-one-dimensional " or `` weakly two-dimensional " system, serves as an intermediate model between 1D and 2D systems \cite{PhysRevLett.116.140401}. In this study, we investigate the many-body localization properties of the fully frustrated Heisenberg spin-1/2 ladder model (hereafter referred to as `` the frustrated ladder model ") using exact diagonalization. To facilitate a comparison with the Heisenberg spin-1/2 single-chain model with next-nearest-neighbor hopping interaction (hereafter referred to as `` the single-chain model "), we also consider next-nearest-neighbor hopping interaction along the leg direction in our frustrated ladder model.

The rest of this article is structured as follows. In Sec. \ref{sec2}, we introduce our model, including its parameters and the study methods. In Sec. \ref{sec3}, we compare the spectrum and eigenstate properties of the two models and approximately determine the critical points. Moreover, we observe the existence of a many-body mobility edge in the frustrated ladder model. Sec. \ref{sec4} shows the dynamics of the frustrated ladder model, encompassing the time evolution of entanglement entropy, fidelity, and magnetization for initial states. In Sec. \ref{sec5}, we perform a finite-size scaling analysis based on two types of MBL transitions. Finally, Sec. \ref{sec6} presents the conclusion.
\section{\label{sec2}model}
	\paragraph{\label{sec:level2}The single-chain model}
	The Hamiltonian of the Heisenberg spin-1/2 single-chain model with next-nearest-neighbor hopping interactions can be expressed as follows:
	\begin{align}
		H_{1} &= \sum_{i=1}^{N-1} J_{1} \left(S_{i}^{x}S_{i+1}^{x}+S_{i}^{y}S_{i+1}^{y}+S_{i}^{z}S_{i+1}^{z}\right) \nonumber \\
		\quad&+ \sum_{i=1}^{N-2} J_{2} \left(S_{i}^{x}S_{i+2}^{x}+S_{i}^{y}S_{i+2}^{y}\right) + W\sum_{i=1}^{N} h_{i}S_{i}^{z}.
  \label{eq1}
	\end{align}

\begin{figure}[t]
		\centering
		\includegraphics[width=1\columnwidth]{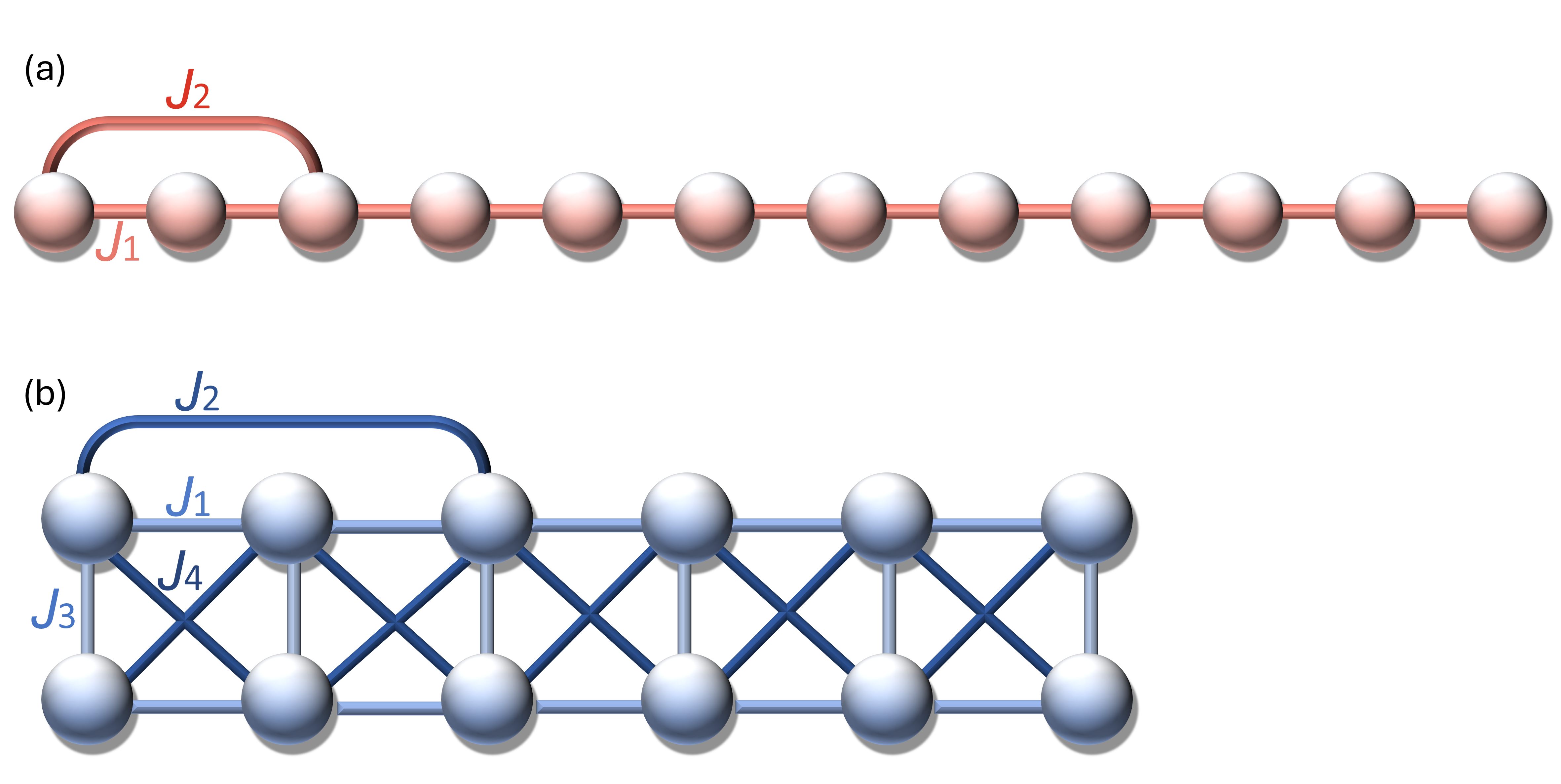}
		\caption{Schematics of our models. (a) The single-chain model. (b) The frustrated ladder model. $J_{1}$ and $J_{2}$ are the nearest-neighbor and next-nearest-neighbor coupling constants along the leg direction, while $J_{3}$ and $J_{4}$ are the nearest-neighbor and next-nearest-neighbor coupling constants along the ladder direction. }
		\label{fig1}
	\end{figure}

	\paragraph{\label{sec:level3}The frustrated ladder model}
	The Hamiltonian of the fully frustrated Heisenberg spin-1/2 ladder model with next-nearest neighbor hopping interaction can be written as follows:
	\begin{align}
	&H_{2}= J_{1} H_{\mid \mid } +J_{2} H_{\mid \mid \mid } +J_{3}H_{\perp} +J_{4} H_{\times }+H_{z},
   \label{eq2}
       \end{align}
where
\begin{align}
	H_{\mid \mid }&=\sum_{j=1,2}\sum_{i=1}^{M-1} \left ( S_{j,i}^{x} S_{j,i+1}^{x}+S_{j,i}^{y} S_{j,i+1}^{y}+S_{j,i}^{z} S_{j,i+1}^{z}  \right ),\nonumber\\
	H_{\mid \mid\mid  }&=\sum_{j=1,2}\sum_{i=1}^{M-2} \left ( S_{j,i}^{x} S_{j,i+2}^{x}+S_{j,i}^{y} S_{j,i+2}^{y}  \right ),\nonumber\\
	H_{\bot } &=\sum_{i=1}^{M} \left ( S_{1,i}^{x} S_{2,i}^{x}+S_{1,i}^{y} S_{2,i}^{y}+S_{1,i}^{z} S_{2,i}^{z}  \right ), \\
	H_{\times  } &=\sum_{i=1}^{M-1} ( S_{1,i}^{x} S_{2,i+1}^{x}+S_{1,i}^{y} S_{2,i+1}^{y})\nonumber\\
        &+\sum_{i=1}^{M-1}(S_{1,i+1}^{x} S_{2,i}^{x}+S_{1,i+1}^{y} S_{2,i}^{y} ), \nonumber\\
	H_{z } &=W\sum_{i=1}^{M} h_{i} \left ( S_{1,i}^{z}+S_{2,i}^{z}\right )\nonumber, 
\end{align}
	where $S_{i}^{\left \{ x/y/z \right \} } $denotes the spin components in the $x$, $y$, and $z$ directions of a spin-1/2 system located at site $i$. We set $J_{1}$ = $J_{2}$ = $J_{3}$ = $J_{4}$ = 1, which corresponds to the fully frustrated ladder model as described in Ref. \cite{10.21468/SciPostPhys.11.4.074}. Simultaneously, we consider $J_{1}$ and $J_{2}$ as the nearest-neighbor and next-nearest-neighbor coupling constants along the leg direction, while $J_{3}$ and $J_{4}$ along the ladder direction, as shown in Fig. \ref{fig1}. $h_{i} $ is a random magnetic field with a uniform and independent distribution $  [ -1,1] $, and $W$ is the strength of the random field. 
	In this work, we set $N$ = $2M$, $N $ represents the number of lattice sites, and $M$ denotes the length of the ladder. We investigate the properties of our models in Eq. (\ref{eq1}) and Eq. (\ref{eq2}) with a size of $N$ = 10,12,...,18 by exact diagonalization within the $S_{ \rm tot}^{z}$ = 0 sector.
 
\section{\label{sec3} spectrum and eigenstate }
In this section, we analyze several important observables that indicate the occurrence of the MBL transition and compare the frustrated ladder model with the single-chain model.
    \paragraph{the adjacent gap ratio }
$\bar{r} $ is a common tool based on the random matrix theory \cite{PhysRevB.75.155111,PhysRevLett.110.084101} for spectrum analysis, used to distinguish the localized phase from the thermalized phase and determine the critical point. The adjacent gap ratio \cite{PhysRevB.92.195153,PhysRevB.91.081103,PhysRevB.75.155111} is defined as
	\begin{equation}
	r_{n}\equiv \rm{min}( s_{n},s_{n+1}) /\rm{max}( s_{n},s_{n+1}) ,
 \end{equation}
   where $ s_{n}=E_{n}-E_{n-1}$. It is important to note that the spectrum is ordered $E_{n}>E_{n-1}$, to determine the critical point. 
As in previous studies \cite{PhysRevB.92.195153,PhysRevB.98.094202}, the target energy density is defined as
   	\begin{align}
   \varepsilon =\frac{E-E_{ \rm min} }{E_{ \rm max}-E_{ \rm min}},
   	\end{align}
    where $E_{ \rm max}(E_{ \rm min})$ is the maximum (minimum) value in the energy spectrum $\left \{  E\right \} $. In our study, we choose the eigenenergies in the window of $ \varepsilon$ = 0.5 $\pm$ 0.01, which is at the middle of the energy spectrum \cite{PhysRevB.92.195153,PhysRevB.100.134504}. In the thermalized phase, the system demonstrates a Wigner-Dyson (WD) distribution with $\left \langle r \right \rangle _{ \rm W} $ = 0.5307. This is caused by the overlap of energy eigenstates, which results in repulsion between energy levels. On the contrary, it conforms to a Poisson distribution with $\left \langle r \right \rangle _{ \rm P}$ = 0.3863 in the localized phase, where there is no level repulsion \cite{WOS:000229464700001}.

     \begin{figure*}[t]
    	\centering
    	\includegraphics[width=2.05\columnwidth]{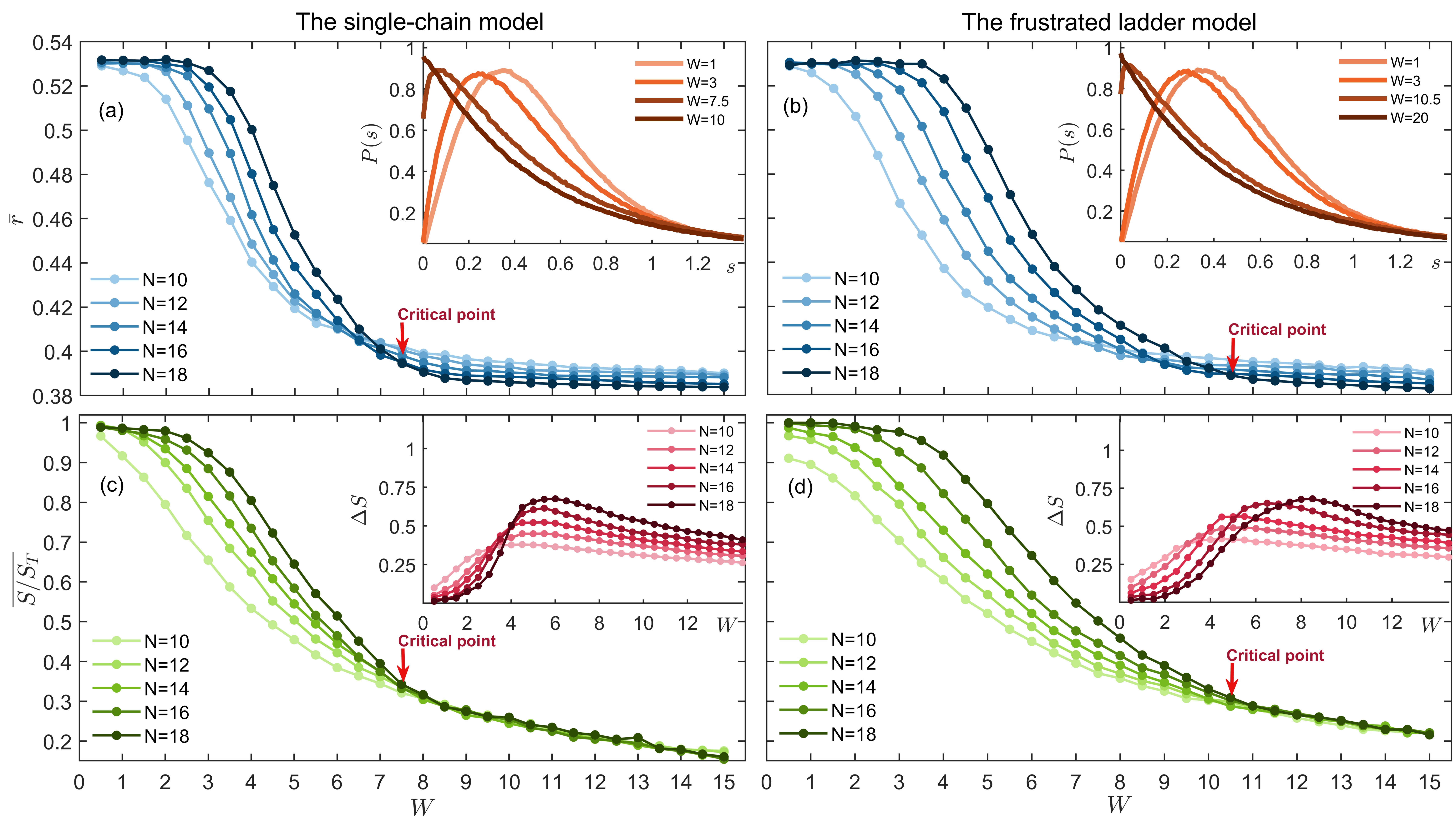}
    	\caption{ Spectrum and eigenstate properties of our two models. [$(a)$ and $(b)$] show the adjacent gap ratio $\bar{r}$ for different $N$ as a function of $W$ for the single-chain model and the frustrated ladder model. They both illustrate the variation of $\bar{r}  $ from $\left \langle r \right \rangle _{ \rm W}$ to $\left \langle r \right \rangle _{ \rm P}$ as $W$ increases in the two models. The critical points are $w _{1} \sim 7.5\pm 0.5$ and $w _{2} \sim10.5\pm 0.5$ (indicated by the red arrows). The insets of [(a) and (b)] depict the level distribution $P(s)$ for different $W$ as a function of $s$. They display the transition of $P(s)$ from the WD distribution to the Poisson distribution as $W$ increases for both models when considering system size $N$ = 16. [(c) and (d)] display $\overline{S/S_{T}}$ for different $N$ as a function of $W$. They exhibit the transition of $ \overline{S/S_{T}} $ from volume-law entanglement in the thermalized phase to area-law entanglement in the localized phase. Notably, the estimated critical points (indicated by the red arrows) align with the values obtained from $\bar{r}$. The insets of [(c) and (d)] show the variance $\bigtriangleup S$ for different $N$ as a function of $W$. 
    	}
    	\label{fig2}
    \end{figure*}

 \paragraph{level-spacing distribution}
 $P(s)$ is a useful tool for understanding the short-range spectral correlations in a system during its long-term evolution. Specifically, $P(s) ds$ stands for the probability that $s_{n} $ is in $(s,s+ds)$. $P(s)$ obeys the WD distribution $P(s)$ = $\frac{\pi s}{2} e^{-\frac{\pi s^{2} }{4} } $ in the thermalized phase and the Poisson distribution $P(s)$ = $e^{ -s }$ in the localized phase \cite{PhysRevB.98.094202,PhysRevB.94.144201}. However, near the critical point, the localization and thermalization features of the system begin to mix, and there are significant quantum fluctuations presented in the system.  As a consequence, $P(s)$ exhibits intermediate behavior between the two distributions.

    \paragraph{entanglement entropy}
   $S$ is a valuable tool for providing crucial information about the entanglement structure and quantum coherence of the system \cite{PhysRevX.7.021013,PhysRevResearch.2.033154}. We divide the system into two parts, $A$ and $B$, with equal lattice size $M$. The entanglement entropy of subsystem $A$ can be expressed as follows:
   	\begin{align}
    S=- { \rm Tr} (\rho _{A} { \rm ln}\rho _{A}),
    	\end{align}
     where $\rho_{A}$ = ${ \rm Tr}_{B} \rho $ = $ { \rm Tr}_{B}| \psi  \rangle\langle\psi| $. In the thermalized phase, the system obeys the volume law, while in the localized phase, it adheres to the area law \cite{PhysRevB.91.081103}.

    \paragraph{the variance of the entanglement entropy} $  \bigtriangleup S $ is defined as 
    \begin{align}
    \left ( \bigtriangleup S \right ) ^{2} = \left \langle S^{2} \right \rangle -\left \langle S \right \rangle^{2}. 
    	\end{align}
 This well-defined quantity serves as an excellent indicator of phase transitions \cite{PhysRevB.91.081103,PhysRevResearch.2.033154}. The variance reaches its maximum value near the critical point.
    
We now move to the results analysis. We average the gap ratio $\bar{r}  $ = $\left \langle r_{n}  \right \rangle$ over eigenvalues included in $\varepsilon \in $ [0.49, 0.51] from each disorder realization and 1000 to 20000 disorder realizations depending on $N$ for each $W$. In Figs. \ref{fig2}\textcolor{blue}{[(a) and (b)]}, we observe that the curves for different $N$ approximately cross at the critical point $w _{1} \sim 7.5\pm 0.5$ (The value is slightly larger than that in Ref. \cite{PhysRevX.7.021013} because we choose a smaller energy window) for the single-chain model, and $w _{2} \sim 10.5\pm 0.5$ for the frustrated ladder model. This implies that the frustrated ladder model requires larger $W$ to undergo the MBL transition. Moreover, we notice that as $N$ increases, the critical points shift towards larger $W$, and this trend is more pronounced in the frustrated ladder model due to the broader distribution of the curves in Fig. \ref{fig2}\textcolor{blue}{(b)}. We can attribute these phenomena to the increased complexity and stronger interactions present in the frustrated ladder model.

The insets of Figs. \ref{fig2}\textcolor{blue}{[(a) and (b)]} demonstrate that for small and large values of $W$, the $P( s ) $ of our two models follow the WD distribution and the Poisson distribution, respectively, which are characteristic of the thermalized phase and localized phase \cite{PhysRevB.98.094202}. However, the insets depict an intermediate behavior between these two distributions near the critical points. Although the $P(s)$ distribution is closer to the Poisson distribution (which can be attributed to finite size effects, similar to Ref. \cite{PhysRevB.94.144201}), it also exhibits a level repulsion for small energy separations
s $\ll$ 0.1.
    
 Figs. \ref{fig2}\textcolor{blue}{[(c) and (d)]} illustrate $ \overline{S/S_{T}} $ as a function of $W$ for different system sizes $N$, where $S_{T}$ = 0.5 ($N$ln2 - 1) is the Page value for random pure states \cite{PhysRevResearch.2.033154,PhysRevX.7.021013}. The number of disorder realizations is 500 to 10000  for $N$ = 18 to 10. With $W$ increase, we observe a transition in  $\overline{S/S_{T}} $ going from 1 in the ETH phase characterized by volume-law \cite{PhysRevB.85.094417} entanglement to 0 in the MBL phase characterized by area-law entanglement. Moreover, the crossings of the curves for the two models align closely with the critical points derived from $\bar{r}$. Additionally, it is worth mentioning that the size effect of entanglement entropy is not as pronounced as that of the adjacent gap ratio.
 
 In addition, as shown in the insets of  Figs. \ref{fig2}\textcolor{blue}{[(c) and (d)]}, when $W\to 0$, $\bigtriangleup S\to 0$, this is consistent with the ETH. Near the critical point, $\bigtriangleup S$ exhibits its maximum fluctuations. The peak values of the curves are slightly smaller than the previously critical points $w_{1}(w_{2})$, thus serving as a lower bound for the occurrence of the phase transition, which is consistent with previous studies in one-dimensional system \cite{Bahri2015} and Heisenberg ladder model \cite{PhysRevB.92.195153}.

 \begin{figure}[t]
    	\centering
    	\includegraphics[width=1\columnwidth]{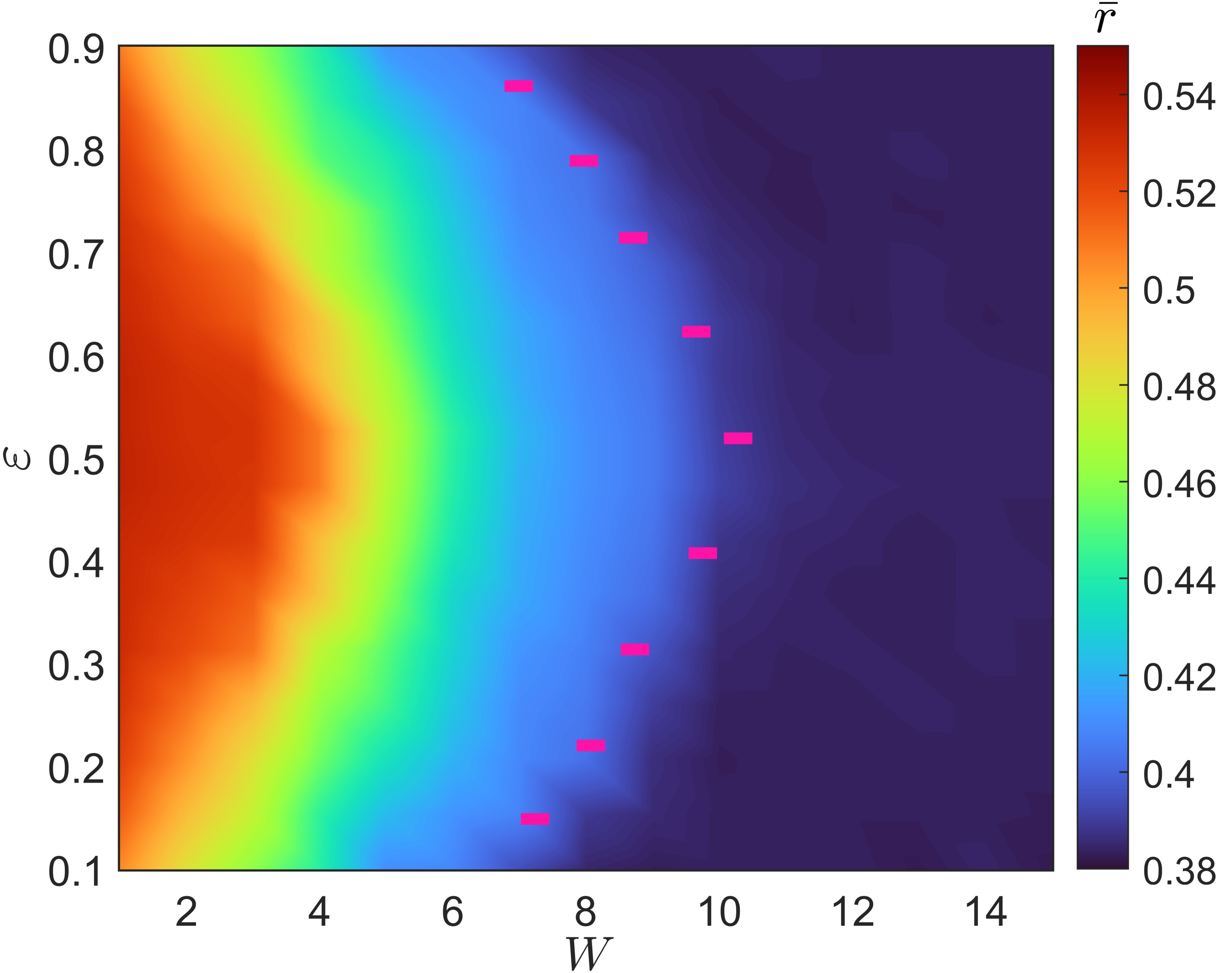}
    
    	\caption{ Disorder strength ($W$)—Energy density ($\varepsilon$) phase diagram of the frustrated ladder model for $N$ = 16. The color bar represents the $\bar{r}$ value ranging from 0.3863 to 0.5307. The pink rectangle mark corresponds to a visual estimate of the boundary between the WD distribution and the Poisson distribution of $\bar{r}$. The “D-shape” structure in the phase diagram exhibits the well-known mobility edge. }
    	 \label{fig3}
\end{figure}

 Furthermore, in Fig. \ref{fig3}, we plot the adjacent gap ratio  $\bar{r}$ as a function of $W$ and $\varepsilon$ for $N$ = 16. To achieve this, we sample energies only in $\varepsilon\in$ [0.1,0.9], because the density of states is particularly low at the edge of the energy spectrum. We partition $\varepsilon$ into 16 equal segments, resulting in a value of $\bigtriangleup \varepsilon$ = 0.05. We then perform 3000 disorder realizations for each $\bigtriangleup \varepsilon$ at different $W$. With $W$ increase, the phase diagram shows a continuous transition from the thermalized phase (red region) to the localized phase (dark blue region). Moreover, the pink rectangle mark corresponds to a visual estimate of the boundary between the WD distribution and the Poisson distribution of $\bar{r}$.

Specifically, at small $W$, the eigenstates within intermediate energy densities are found to be thermalized, while the eigenstates for edge energy densities exhibit localization. 
The “D-shaped” structure in the phase diagram of the frustrated ladder model is consistent with observations in one-dimensional models \cite{PhysRevB.91.081103}, ladder models, and triangular models \cite{PhysRevB.98.094202}. However, its boundary corresponds to larger $W$. 

 \section{\label{sec4}Dynamics}
 In the previous sections, we compared the properties of the two models. Here, our focus shifts to the dynamic evolution of the frustrated ladder model, which represents an additional significant approach for studying MBL transition. 
 We choose the $\rm{N\acute{e}el}$ state as the initial state $\left | \psi (0)  \right \rangle $ for numerical simulations in our model. We average 300 disorder realizations on the system with $N$ = 14 at different $W$. Additionally, we analyze the size effects for different $N$ with the fixed disorder strength $W$ = 3. The number of disorder realizations is 100 to 5000 for $N$ = 16 to 10.

    \paragraph{entanglement entropy}
   The growth behavior of entanglement entropy $S(t)$ serves as a significant criterion for distinguishing between the thermalized phase and the localized phase \cite{PhysRevB.100.134504,PhysRevB.85.094417,PhysRevB.107.144201}. We divide the system into two parts, $A$ and $B$, with equal lattice size $M$. The evolution of entanglement entropy is defined by setting
    	\begin{align}
    &S(t) =-{ \rm Tr}\left [  \rho _{A}(t) { \rm ln}\rho _{A}(t)  \right ], 
    \end{align}
  
    where $\rho _{A}(t)$ = ${ \rm Tr}_{B} \rho(t)$ = ${ \rm Tr}_{B} |\psi(t)\rangle\langle\psi(t)|$ is the reduced density matrix of the system after tracing out subsystem $B$, while $\left|\psi(t)\right\rangle$ = $e^{-iHt}\left|\psi(0)\right\rangle$ is the state obtained after the evolution of the initial state $\left|\psi(0)\right\rangle$ for time $t$.
    
\begin{figure*}[t]
   	\centering
   	\includegraphics[width=2.1\columnwidth]{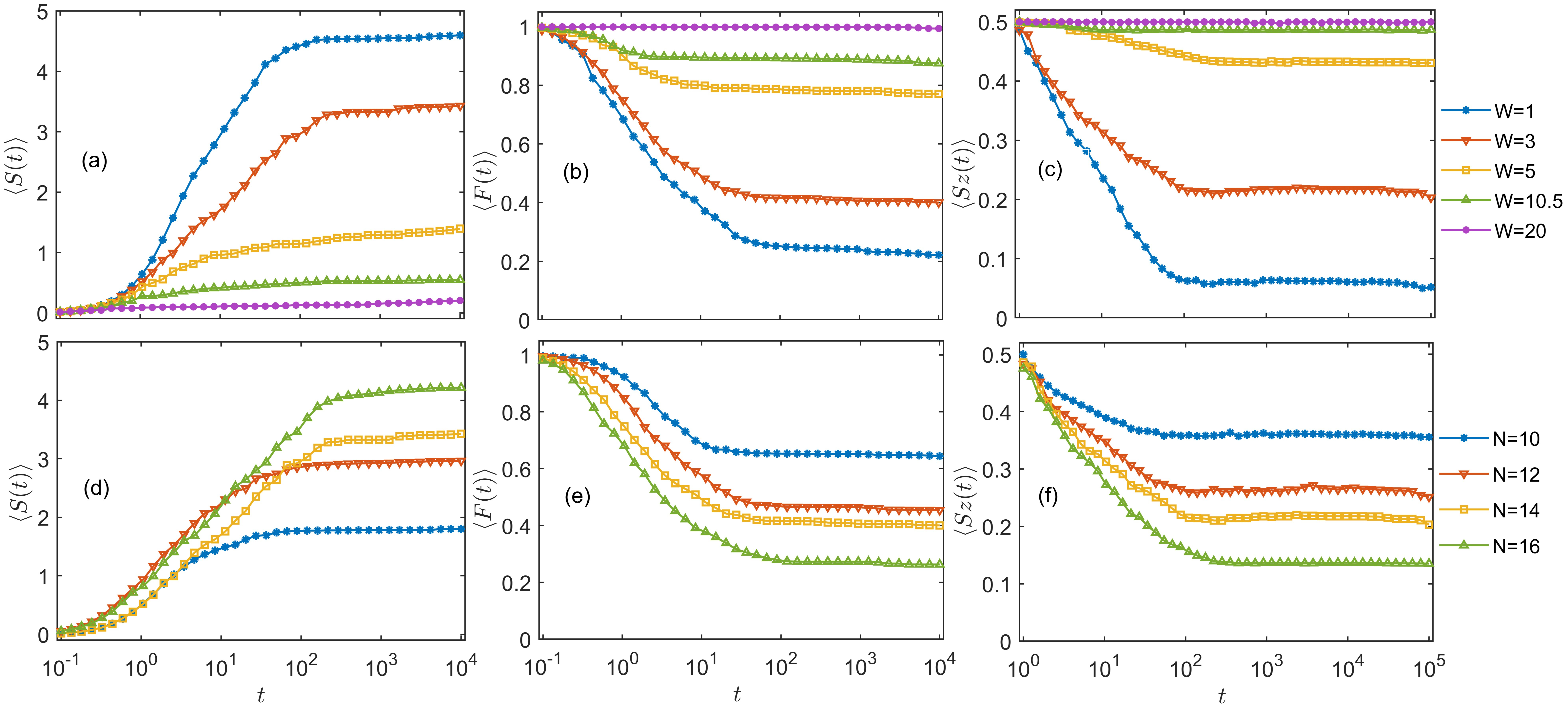}
   	\caption{The time evolution of quantities in the frustrated ladder model.
[(a), (b), and (c)] depict the distinct dynamical evolution behaviors of entanglement entropy, fidelity, and magnetization, for different $W$ at $N$ = 14. At small $W$, (a) displays the rapid power-law growth of entanglement entropy, (b) illustrates the loss of initial information, and (c) depicts the behavior of free magnetization, which are characteristics of the ETH phase. However, at large $W$, (a) depicts the slow logarithmic growth of entanglement entropy, (b) shows the high fidelity of initial information, and (c) displays the presence of initial magnetization, which are characteristics of the MBL phase. Additionally, [(d), (e), and (f)], show the size effects for $N$ = 10 to 16 at $W$ = 3. For the same disorder strength $W$, as $N$ decreases, the properties of the system become more similar to those of the MBL phase. }
   	\label{fig4}
   \end{figure*}

    Fig. \ref{fig4}\textcolor{blue}{(a)} illustrates the evolution of entanglement entropy with time for $N$ = 14. In the thermalized phase ($W$ = 1), a rapid power-law growth of the entanglement entropy is evident, which quickly saturates to a large value within a short period of time. This value is close to the random pure state Page value $S_{T}$ = 0.5 ($N$ ln2 - 1)  of the volume law. In contrast, in the deep localized phase ($W$ = 20), the entropy exhibits a distinct slow logarithmic growth (related to the integration of local integrals of motion) and saturates to a small value over a long period of time. It is worth noting that for intermediate disorder strengths ($W$ = 5), the entanglement entropy initially undergoes a short period of power-law growth, where information spreads rapidly. However, unlike in the thermalized phase, the spread of information almost comes to a halt within a short period of time. Instead, it exhibits a long-lasting slow logarithmic growth trend. This is consistent with the behavior observed in one-dimensional systems \cite{PhysRevB.96.075146,PhysRevB.91.081103}.
   
    In  Fig. \ref{fig4}\textcolor{blue}{(d)} we also analyze the finite-size effects for systems with sizes ranging from $N$ = 10 to 16 and $W$ = 3. We find that as $N$ increases, the entanglement entropy saturates to a larger value. More importantly, as $N$ increases, the time for power-law growth of the entanglement entropy increases, while the time for logarithmic growth decreases. This suggests that systems with larger $N$ will localize at larger $W$, which is consistent with the discussion in Fig. \ref{fig2}\textcolor{blue}{(d)} that as $N$ increases, the crossing point shifts to larger $W$.

    \paragraph{fidelity}
$F(t)$ is an important observable in quantum information, used to measure the similarity between the initial state and the evolved state, allowing us to determine the properties of the system. The evolution of fidelity \cite{PhysRevLett.99.100603,PhysRevA.75.032109}is defined as 
     \begin{align}
    F(t) =\rm{Tr}\left [ \rho(t)^{1/2}\rho _{0} \rho(t)^{1/2}\right ] ^{1/2}, 
       \end{align}    
 where $\rho(t)$ = $|\psi(t)\rangle\langle\psi(t)| $ and $\rho(0)$ = $|\psi(0)\rangle \langle\psi(0)|$. The evolved state $\left|\psi(t)\right\rangle$ = $e^{-iHt}\left|\psi(0)\right\rangle$.
 
 In Fig. \ref{fig4}\textcolor{blue}{(b)}, we observe that at small disorder strength ($W$ = 1), the fidelity stabilizes at a relatively small value after long-time evolution, indicating that the initial state information is almost completely lost, consistent with the ETH phase. On the other hand, at large disorder strength ($W$ = 20), the fidelity remains constant with time, indicating that the initial state information is fully preserved, in line with the MBL phase. Additionally, we find that as $W$ increases, the preservation of initial state information becomes higher. 
 
 Furthermore, in Fig. \ref{fig4}\textcolor{blue}{(e)}, we also analyze the influence of $N$ on the system properties at $W$ = 3. We find that for the same $W$, as $N$ increases from 10 to 16, the fidelity takes a longer time to stabilize, and the stability value is smaller. We believe that in smaller systems, there are fewer interactions between spins, resulting in fewer paths required for information transmission. This makes the system more sensitive to random perturbations and thus enables faster information propagation and retention. 
  \begin{figure*}[t]
 	\centering
 	\includegraphics[width=2.05\columnwidth]{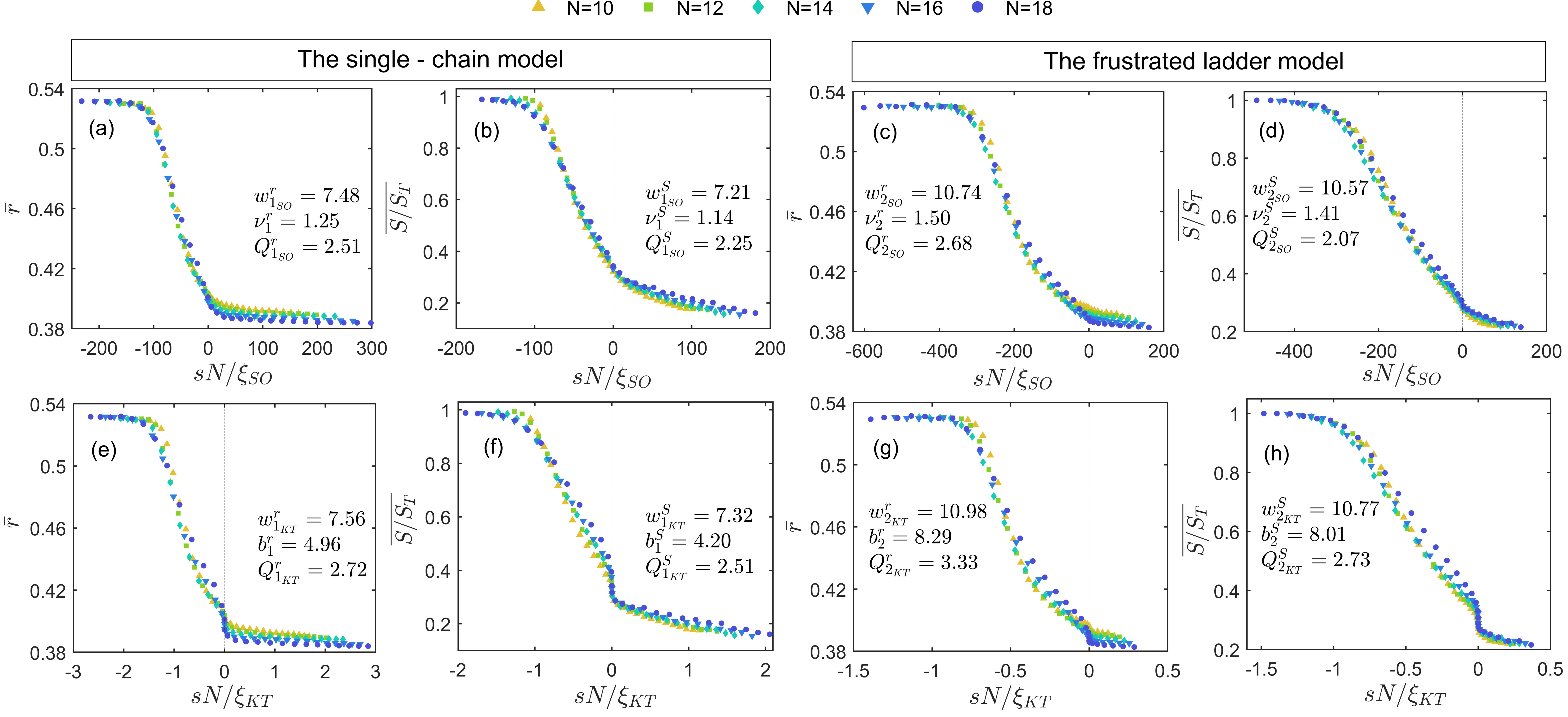}
 	\caption{\textbf{Upper panels:} the finite-size scaling of quantities as a function of $sN/\xi _{SO}$, where $\xi_{SO}$ is the correlation strength in Eq. (\ref{eq.11}) and $s$ = ${\rm{sgn}}[W-w]$. [(a) and (b)] $\bar{r}$ and $\overline{S/S_{T} }$ for the single-chain model, [(c) and (d)] $\bar{r}$ and $\overline{S/S_{T} }$ for the frustrated ladder model. \textbf{Lower panels:} the finite-size scaling of quantities as a function of $sN/\xi _{KT}$, where $\xi_{KT}$ is the correlation strength in Eq. (\ref{eq.12}). [(e) and (f)] $\bar{r}$ and $\overline{S/S_{T} }$ for the single-chain model, [(g) and (h)] $\bar{r}$ and $\overline{S/S_{T} }$ for the frustrated ladder model. In each panel, we provide the parameters obtained by minimizing $Q$ for optimizing the best data collapse.}
 	\label{fig5}
 \end{figure*}
    \paragraph{magnetization}
    $M_{I} (t)$ is an important property used to characterize MBL transition. \cite{PhysRevB.103.L100202,PhysRevB.91.140202,PhysRevLett.114.140401}. By calculating the average magnetization of a given lattice site overall eigenstates, we can study its time evolution and analyze system properties. The definition of the expectation value of magnetization at the site $I$ is provided by 
     \begin{equation}
   M_{I} (t) =\langle\psi(t)|S_{I}^{z}| \psi(t)\rangle, 
      \end{equation}
   where $\left|\psi(t)\right\rangle$ = $e^{-iHt}\left|\psi(0)\right\rangle$, and here we let $I$ = 1.
   
   As shown in Fig. \ref{fig4}\textcolor{blue}{(c)}, when the disorder strength is small ($W$ = 1), the long-term evolution of the magnetization of the site $I$ tends to a very small value. This corresponds to the prediction of the ETH for an effectively free spin, indicating the loss of initial information. However, when the disorder strength is large ($W$ = 20), the average magnetization stabilizes at 1/2 after long-time evolution, indicating good preservation of the initial state information. The behavior for intermediate disorder strengths $W$ falls between the localized phase and the thermalized phase. The reason behind these phenomena is the spins in the thermalized phase are correlated over long distances in space, which causes the magnetization to decay exponentially as time progresses. Conversely, in the localized phase, the system shows magnetic localization, where spins are randomly distributed in space and do not exhibit long-range order. As a result, the system only maintains strong spin correlations within a limited range, leading to a more stable evolution of magnetization over time. Furthermore, we also observe that as $W$ decreases, the decay of magnetization is faster. The reason behind this is that when the disorder strength is small, the spins on individual lattice sites are less affected, leading to a faster convergence of their evolution towards a stable state.
   
   In Fig. \ref{fig4}\textcolor{blue}{(f)}, we also analyze the size effect for $W$ = 3, which shows consistency with the dynamic properties of fidelity. Specifically, as $N$ increases, the average magnetization decays faster and stabilizes at a smaller value after a longer period. Based on our analysis, we can conclude that in smaller systems, the limited range of interactions between spins leads to a stronger localization effect. Additionally, in smaller systems, the energy level spacing is smaller and the degeneracy of energy levels is higher. This affects the thermal fluctuation behavior of the spins, which can cause the spins in smaller systems to remain consistent for longer periods in more states, resulting in a slower decay rate for the average magnetization.
\section{\label{sec5} finite-size scaling analysis}
Now, let's compare the finite-size scaling properties of our two models. Currently, the description of the MBL transition is mainly based on two hypotheses, both of which suggest that the length scale $\xi$ diverges near the critical point. Most literature suggests that MBL follows a continuous second-order phase transition \cite{PhysRevX.7.021013,
PhysRevB.94.144201,PhysRevLett.119.075702,PhysRevLett.113.107204}, with its power-law divergent correlation strength 
\begin{equation}
\xi _{SO}\sim {|W-w |^{-\nu } },
\label{eq.11}
\end{equation}
where $w $ is the critical point and $\nu$ is the universal critical exponent. In this hypothesis, the famous Harris criterion states that $\nu \ge 2/d$ and $d$ is the dimension of the system. Nevertheless, most numerical simulations based on one-dimensional systems show $\nu \sim 1$ \cite{PhysRevB.91.081103,PhysRevLett.119.075702}, which contradicts the Harris bound $\nu \ge 2$. Two explanations have been put forward to account for this discrepancy. One is that the simulated system sizes are insufficient to emulate the thermodynamic limit. Alternatively, the hypothesis proposing a continuous second-order phase transition for MBL is valid.

 \begin{figure*}[t]
 	\centering
 	\includegraphics[width=2.08\columnwidth]{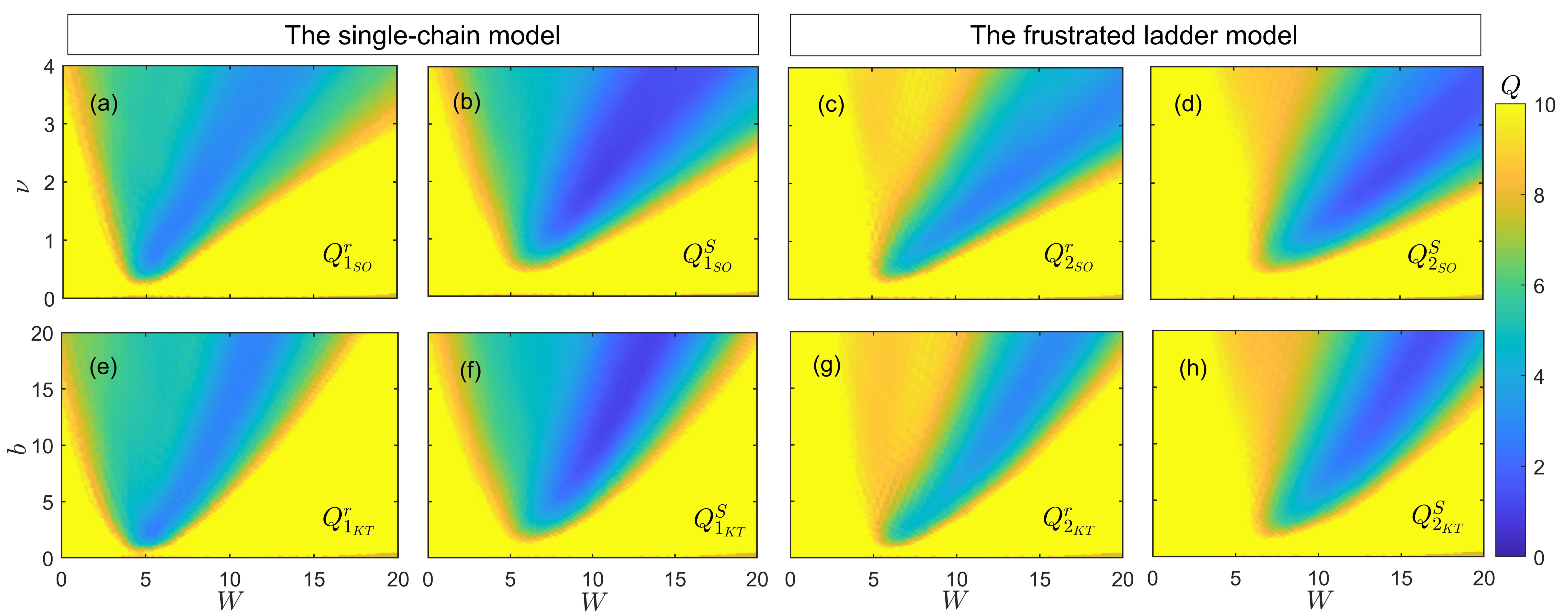}
 	\caption{Disorder strength $W$—Universal critical exponent $\nu$ (Nonuniversal parameter $b$) phase diagram. The color bar represents the value of quality function $Q$. \textbf{Upper panels:} The $Q$ values of quantities as a function of $W$ and $\nu$ with the correlation strength $\xi _{SO}$. [(a) and (b)] $\bar{r}$ and $\overline{S/S_{T} }$ for the single-chain model. [(c) and (d)] $\bar{r}$ and $\overline{S/S_{T} }$ for the frustrated ladder model. \textbf{Lower panels:} The $Q$ values of quantities as a function of $W$ and $b$ with the correlation strength $\xi _{KT}$. [(e) and (f)] $\bar{r}$ and $\overline{S/S_{T} }$ for the single-chain model, [(g) and (h)] $\bar{r}$ and $\overline{S/S_{T} }$ for the frustrated ladder model.}
 	\label{fig6}
 \end{figure*}

Recently, an improved real-space renormalization group (RG) scheme has proposed a Kosterlitz-Thouless (KT) type MBL transition \cite{PhysRevB.102.125134,
PhysRevB.99.224205,
PhysRevLett.122.040601,
PhysRevB.99.094205}, in which the correlation length 
\begin{equation}
\xi _{KT}\sim {\rm{exp}}({{b|W-w |^{-0.5  }}}) ,
\label{eq.12}
\end{equation}
where $w $ is the critical point and $b$ is a nonuniversal parameter. Although several articles have compared the MBL transition with the two length scales $\xi _{SO}$ and $\xi _{KT}$, no conclusive answer has been provided regarding which one is more suitable for finite-size scaling. Some have concluded that $\xi _{SO}$ provides better data collapse than $\xi _{KT}$ \cite{PhysRevB.104.214201}, while others have suggested the opposite \cite{PhysRevE.102.062144,PhysRevB.102.064207}. Therefore, motivated by these results we perform finite-size scaling on our two models with the two correlation lengths $\xi _{SO}$ and $\xi _{KT}$.

To obtain more precise estimates of $w $ and other important parameters, we present a finite-size scaling analysis on the curves shown in Fig. \ref{fig2}.
In the upper panels of Fig. \ref{fig5}, we plot $\bar{r}$ and $\overline{S/S_{T} }$ as a function of $sN/\xi _{SO}$, where $\xi_{SO}$ is the correlation strength in Eq. (\ref{eq.11}), while in the lower panels, we plot them as a function of $sN/\xi _{KT}$, where $\xi_{KT}$ is the correlation strength in Eq. (\ref{eq.12}) and $s$ = ${\rm{sgn}}[W-w]$ for our two models. By varying the values of  $w $, $\nu $ ($b$),   the curves for different $N$ collapse on each other. We can achieve the best possible data collapse by minimizing a quality function 

\begin{equation}
Q=\frac{\sum_{j=1}^{N_{p}-1 }|Y_{j+1} -Y_{j}| }{{\rm{max}}\left \{ Y_{j}  \right \} -{\rm{min}}\left \{ Y_{j}  \right \}} -1,
\label{eq.13}
\end{equation}
given in Refs. \cite{PhysRevB.104.235112,PhysRevE.102.062144,PhysRevB.104.214201,PhysRevB.102.064207,PhysRevB.107.045108}. In Eq. (\ref{eq.13}), $Y$ = $\left \{ Y_{j}  \right \}$ represents a quantity with $Np$ values at various $N$ and $W$, and it is sorted in nondecreasing order based on the values of $sN/\xi_{SO}$ or $sN/\xi_{KT}$. 
Under the best data collapse, where all curves collapse onto a single curve, $Q$ = 0, otherwise $Q$ $>$ 0. In our analysis, we select the parameters that minimize $Q$. 

As shown in Fig. \ref{fig5}, the values of $w$ obtained from the best scaling solution approximately consist with the crossing points of curves for different $N$ in Fig. \ref{fig2} for our two models. Furthermore, the $\nu$ values obtained from our two models both violate the Harris bound, consistent with previous studies \cite{PhysRevB.91.081103,PhysRevLett.119.075702} which prevents us from using the universal critical exponent $\nu$ to classify the MBL transition of our two models as done in Ref. \cite{PhysRevResearch.2.013163}. More importantly, we obtain smaller values of $Q$ from the scaling solution quantities as a function of $sN/\xi_{SO}$ than from $sN/\xi_{KT}$ for both models. We also observe that $\overline{S/S_{T} }$ produces smaller $Q$ compared with $\bar{r}$ for both models and both scaling solutions. To further support our claims, we plot $Q$ as a function of $w$ and $\nu (b)$ in Fig. \ref{fig6}. We also observed a smaller $Q$ (more pronounced dark blue) for $\overline{S/S_{T} }$ as a function of $sN/\xi$ compared with $\bar{r}$. This is consistent with the smaller size effect of entanglement entropy shown in Fig. \ref{fig2}. Based on the comparisons, we conclude that $\xi_{SO}$ yields a better scaling solution than $\xi_{KT}$, as evidenced by the smaller $Q$, and this is more pronounced in the frustrated ladder model ($\bigtriangleup Q$ = $Q_{2_{KT}}-Q_{2_{SO}}\approx 0.7$) compared with the single-chain model ($\bigtriangleup Q$ = $Q_{{1_{KT}}}-Q_{2_{SO}}\approx 0.2$).

\section{\label{sec6}conclusion}
In this article, the MBL properties of the fully frustrated ladder model (with the next-nearest neighbor hopping interaction along the leg direction) are investigated and compared with the Heisenberg spin-1/2 single-chain model with the next-nearest neighbor hopping interaction. Similar MBL properties were found in these two models. However, there are differences, as the fully-frustrated ladder model requires stronger disorder ($w _{2} \sim$ 10.5 ± 0.5) to undergo the MBL transition and exhibits more pronounced size effects due to its stronger and more complex interactions.
Additionally, intriguing phenomena were observed in the fully-frustrated ladder model. In the thermalized phase, the power-law growth of entanglement entropy, loss of initial information, and the emergence of effective free-magnetization were observed. Conversely, in the localized phase, logarithmic growth of entanglement entropy, highly preserved initial information, and magnetic localization phenomenon were observed.
Furthermore, the finite-size scaling properties of our two models are studied. Our analysis shows that the continuous second-order phase transition interpretation of the MBL transition provides a better scaling solution than the Kosterlitz-Thouless (KT) type transition for our two models, and this effect is even more pronounced in the frustrated ladder model due to its larger $\bigtriangleup Q$. 
	
	\section*{Acknowledgement}

This work was supported by the Plan for Scientific and Technological Development of Jilin Province (No. 20230101018JC).

\nocite{*}

\bibliography{references}

\end{document}